\documentclass[12pt,preprint]{aastex}

\newcommand\Msolar{\mbox{$M_\odot$ }}

\shorttitle{EVLA Imaging Line Survey of AGB Stars}
\shortauthors{Claussen et al.}

\begin{document}


\title{A Pilot Imaging Line Survey of RW~LMi and IK~Tau Using the \\
Expanded Very Large Array}


\author{M. J Claussen, L. O. Sjouwerman, M. P. Rupen }
\affil{National Radio Astronomy Observatory, Socorro, New Mexico 87801}
\author{H. Olofsson, F. L. Sch\"oier\altaffilmark{1}, P. Bergman}
\affil{Onsala Space Observatory, Dept. of Radio and Space Science, Chalmers University
of Technology, 43992 Onsala, Sweden}
\altaffiltext{1}{deceased}
\author{G. R. Knapp\altaffilmark{2}}
\affil{Department of Astrophysical Sciences, Princeton University, Princeton, NJ 08544}
\altaffiltext{2}{Visiting Astronomer, NRAO}


\begin{abstract}
We report on a pilot imaging line survey (36.0 - 37.0 GHz, with $\sim$1 km s$^{-1}$
spectral channels) with the Expanded Very Large Array for two 
asymptotic giant branch stars, RW~LMi ($=$ CIT6, which has a carbon-rich circumstellar 
envelope) and IK~Tau ($=$ NML~Tau, with an oxygen-rich circumstellar envelope).  Radio continuum 
emission consistent with photospheric emission was detected from both stars.
From RW~LMi we imaged the HC$_3$N (J $=$ 4$\rightarrow$3) emission.
The images show several partial rings of emission; these multiple shells trace the
evolution of the CSE from 400 to 1200 years.  SiS (J $=$ 2$\rightarrow$1) emission was 
detected from both RW~LMi and IK~Tau.  For both stars the SiS emission is centrally 
condensed with the peak line emission coincident with the stellar radio continuum
emission.  In addition, we have detected weak HC$_7$N (J $=$ 32$\rightarrow$31)
emission from RW~LMi.

\end{abstract}


\section{Introduction}

The extensive mass loss which occurs from stars on the asymptotic
giant branch (AGB) leads to the formation of dusty, extended, circumstellar
envelopes (CSE) and the production of many gas-phase molecular species in
the atmospheres and CSE (e.g.\ Olofsson 2008). The CSEs of 
AGB stars eventually contribute their heavy elements and dust grains back to
the interstellar medium (ISM).  Molecules from the CSE may be returned
directly to the ISM or by incorporation into or onto the dust grains.

Mass loss from AGB stars can be as high as 10$^{-4}$ \Msolar yr$^{-1}$ or even
higher (Habing 1996) and is the dominant process at this stage of evolution.
It determines the stellar lifetime on the AGB, the maximum luminosity
reached, the post-AGB evolution of the star, the gas and dust return to the
ISM, and the chemical composition of the returned gas (Bloecker 1995). 
Although the mass loss mechanism is complex and dependent on many
different physical and chemical processes, the reasonably simple geometry
and kinematics of the resulting CSEs make them excellent astrophysical
and astrochemical laboratories.


Interferometric imaging of thermal molecular emission in CSEs have been
mostly confined to the nearby carbon-rich object
IRC+10\arcdeg216, which has a very high mass-loss rate, although there
have been targeted imaging projects of other sources.  If we are
interested in a much better understanding of circumstellar chemistry, we must
observe more typical sources, compare oxygen- and carbon-rich
CSEs, use high-sensitivity interferometry to provide detailed images of the molecular
line distribution, and cover a large number of molecular species transitions.  
To do all these successfully
one line at a time (or even a few) would be prohibitive in terms of
telescope time.  
Thus an interferometric imaging line survey, as being
carried out by the Submillimeter Array (Patel et al. 2011) is
required.

In this short {\it Letter} we report on a project that is 
a pilot for observations of such an imaging line survey using the Expanded
Very Large Array (EVLA; Perley et al. 2011).  We proposed to
observe one each of a carbon-rich CSE (RW~LMi $=$ CIT6), an oxygen-rich CSE 
(IK~Tau $=$ NML~Tau), 
and an S-type AGB star ($\chi$~Cyg).  At the time of observations, the WIDAR
correlator could process a maximum of 1 GHz bandwidth per polarization with 
spectral channels of $\sim$1 km s$^{-1}$ at the Ka receiver band (26.5 --- 40
GHz).  We were unable to observe $\chi$~Cyg because necessary Local Sidereal
Time (LST)  coverage was unavailable.

\section{Observations and Data Reduction}

IK~Tau and RW~LMi were observed with the EVLA in the {\bf C}
configuration during 2010 December and 2011 January. Each target was
observed using four dynamic scheduling blocks (observing runs) of 3.5 
hours, yielding a total of about 7 hours on source integration time each 
with a fixed frequency setting simultaneously covering 36.000 to 36.512 and
36.744 to 37.000 GHz. For each subband of 128 MHz bandwidth, 1024 channels with a
channel separation of 125 kHz ($\sim$ 1 km\,s$^{-1}$) were
produced. Due to limitations during the early commissioning phase, the
full one-GHz range was not available at this resolution, and a
relatively large amount of time was spent on additional calibration.
For IK~Tau, 3C~48 was used for absolute flux calibration, 3C~84 for bandpass and delay
calibration, and J0409+1217 for gain (amplitude and phase) calibration.
For RW~LMi, 3C~286 was used for absolute flux calibration, 3C~273 for bandpass and
delay calibration, and J0958+3224 for gain calibration.

Standard data editing, flux, bandpass and phase calibration were
performed using AIPS and a modified version of the VLA pipeline. The
four data chunks per source were combined using the strongest line by
correcting for the approximate Doppler shift for the day of
observation, although the diurnal variation was ignored for these
short observations. The maximum shifts applied were 5 and 8 channels
for IK~Tau and RW~LMi, respectively. Before imaging all individual
channels using natural weighting and 0.165\arcsec\ pixel size without
cleaning, the data were averaged to 20 second visibilities and
separated by subband, which speeds up the process considerably. 
The continuum emission and spectral channels containing interesting
spectral line features were selected and re-imaged with cleaning.

The synthesized beams from the imaging were $\sim$0.7 x 0.6 arcseconds, 
depending slightly upon the frequency and the source.  The typical rms noise
in a single 125 kHz spectral channel is $\sim$500 $\mu$Jy beam$^{-1}$, and
$\sim$10 $\mu$Jy beam$^{-1}$ in the continuum.

\section{Results}

Radio continuum emission at $\sim$36.5 GHz was detected from both IK~Tau 
(970$\pm$10 $\mu$Jy beam$^{-1}$) and RW~LMi (890$\pm$10 $\mu$Jy beam$^{-1}$).


Line emission was detected from the HC$_3$N (J $=$ 4$\rightarrow$3; rest frequency
36392.332 MHz),  the SiS (J $=$ 2$\rightarrow$1; 36309.627 MHz), 
and the HC$_7$N (J $=$ 32$\rightarrow$31; 36095.546 MHz) transitions 
in the carbon-rich CSE in RW~LMi.  The only spectral line clearly detected
in IK~Tau was the SiS (J $=$ 2$\rightarrow$1 transition, although the total 
frequency coverage included the SO (J, N $=$ (2,3)$\rightarrow$(2,2);
36202.041 MHz) transition, and the OCS (J $=$ 3$\rightarrow$2; 36488.813 MHz) transition.

Figure 1 shows the emission in nine of the 34 velocity images of the HC$_3$N line
from RW~LMi.
These images show striking concentric
rings in the HC$_3$N emission; very clear evidence of multiple spherical shells of emission. 
In the large-scale structure of the ring emission, there are also noticeable asymmetries ---
in the west and northwest near the central velocities, and to the east in more blue-shifted
velocities.
The EVLA data is more sensitive at higher angular and spectral resolution than the spectrally
averaged HC$_3$N (J $=$ 5$\rightarrow$4) data from the VLA at 45.5 GHz presented by
Dinh-V-Trung \& Lim (2009).  
The similarities of the two data sets are striking in terms of
asymmetries; however the higher sensitivity EVLA images show a wealth of detail (e.g.\
the several emission shells) not found in the VLA data.

Figure 2 shows images of the continuum-subtracted, averaged (in velocity) emission 
of the SiS line from both RW~LMi and IK~Tau (all channel images showing emission were 
averaged together after subtracting the continuum in the {\it u-v} data).  As 
compared with the HC$_3$N emission, the SiS emission from RW~LMi is more compact, 
centrally condensed, but slightly elongated in the N-S direction.  For IK~Tau the 
SiS emission is also centrally concentrated toward the star, and is also elongated
rather than circular.

Although we detected the HC$_7$N line in RW~LMi, the emission is quite weak, and further
analysis of this transition will be deferred to another publication.

Figure 3 shows the global emission of the HC$_3$N and SiS lines from RW~LMi.  These spectra 
were made
by summing the emission in the same rectangular region of each spectral channel image.
The channel with the largest emission extent was used to set the rectangular region for
each line. The spectra for both the SiS and HC$_3$N lines are clearly asymmetric with
the red-shifted side being the brightest.

\section{Discussion}

\subsection{Radio Continuum Emission}

At submillimeter and infrared wavelengths the continuum emission from AGB stars is dominated by warm 
circumstellar dust which has a very steep spectral index (Draine 2006 and references therein).  
We compared the detections of the radio continuum 
emission at 36.5 GHz with measurements of other authors at different frequencies.  
Marshall et al. (1992) observed both RW~LMi and IK~Tau at frequencies of 264, 394, and 685 GHz.
They find the derived spectral index from 100 $\mu$m (using {\it IRAS} data) through 
1.1 mm to be $\alpha = -$3.9 (IK~Tau) and $-$3.5 (RW~LMi). 
At centimeter wavelengths emission from dust is small compared to that from the radio
``photosphere'' (Reid \& Menten 1997) i.e.\ blackbody emission from or just outside the
stellar photosphere with a temperature approximately 2000 K, and with a spectral index near $-$2.0.

Radio continuum measurements of RW~LMi at $\sim$3 mm or longer include 
13.3 mJy beam$^{-1}$ at 112 GHz (Neri et al. 1998), 8 mJy  at 90.7 GHz (Lindqvist et al. 2000), 
and 2.4 mJy beam$^{-1}$ at 43.3 GHz (Dinh-V-Trung \& Lim 2009).  A two-point spectral index 
derived from 36.5 GHz to 90.7 GHz and from 36.5 GHz to 112 GHz is consistent and equal to 
$\alpha = $ $-$2.3.  The spectral index of 2.3 is reasonably close to that of thermal blackbody 
radiation.  
Spectral indices calculated using the flux measured at 43.3 GHz to both higher and lower frequencies
are inconsistent and do not fit either photospheric emission or dust emission.  All other existing
continuum data agree with either photospheric or dust emission, and in particular the present EVLA
measurements are likely detecting the photosphere of RW~LMi.

A radio continuum measurement of IK~Tau at 96.7 GHz of 6.1 mJy beam$^{-1}$ was reported by Marvel (2005).
The measurement presented here, combined with Marvel's, produce a two-point spectral index
of $-$2.0.  This is just what is expected from a radio photosphere emitting as a blackbody.

\subsection {Line Emission}
\subsubsection{HC$_3$N Emission in RW~LMi}

We used the AIPS task IRING to azimuthally sum the HC$_3$N emission in RW~LMi in continuous rings 
of a fixed width centered on the continuum, for each velocity channel.  Using this method we then
measured the angular radii of the flux density peaks (rings in Figure 1) in each channel.
Assuming that we then can follow the radius of each ring in different channels we can plot, for each
detected ring, the radius versus the (LSR) velocity for each ring.  Based on this preliminary
analysis, we distinguished four such rings of emission; a more careful examination of the channel
images in conjunction with the ring analysis is certainly required.  The plot of radius versus 
velocity is shown in Figure 4.  The radius in arcseconds is transformed to the radius in AU 
using a distance of 440 pc (Sch\"oier et al. 2002).

If the rings of emission actually correspond to spherical shells of molecular emission around the star, 
the radius of the shell as a function of radial velocity is given by 

\begin{displaymath}
r\left(v - v_{*}\right) = r_0 \sqrt{1 - \left( (v - v_{*}) / v_{exp}\right) ^{2}}
\end{displaymath}
where $v_{*}$ is the stellar radial velocity, $r_{0}$ is the shell radius, and $v_{exp}$
is the constant expansion velocity for a given shell (e.g.\ Bowers et al. 1983).  We fit 
this equation to the data obtained for the four HC$_3$N shells as shown in Figure 4, using a 
simple least-squares method, and using $v_{*}, r_{0}$, and $v_{exp}$ as (bounded) free parameters 
in the fit.  The parameters for each of the shells are listed in Table 1.  For comparison, the
expansion velocity and stellar radial velocity are determined to be 17 km s$^{-1}$ and $-$1.0
km s$^{-1}$ (with respect to the LSR), respectively, based on $^{13}$CO observations 
(Sch\"oier \& Olofsson 2000), and the expansion velocity estimated from the spectra in Figure 3 
is also 17 --- 18 km s$^{-1}$.

Given the expansion 
velocity and the radius, a time can be derived which is the approximate age of the shell.
This age is also listed in Table 1.  The age of shell 3 (from Table 1; blue symbols in Figure 4)
is essentially the same as the age of shell 2 (red symbols in Figure 4) --- this is clearly
because the fit from the data for shell 3 gives a much larger expansion velocity than
that from shell 2.   We speculate that since these two shells have the same age, they may have
a common origin.  If so, this would suggest a rather complex behavior in the mass-loss history
for RW~LMi.

These fitted parameters are only approximate, since it is not always clear how to associate 
the data into rings. (The red symbols in Figure 4 are an example of this.)  Also, the equation 
above (essentially an ellipse) does not fit the blue-shifted side of the data very well: note 
that the data on the blue-shifted side of Figure 4 are quite linear as compared to the data 
on the red-shifted side (which are fit by the velocity equation much better).  Perhaps a better
way of saying this is that, except for shell 1 (which does not have much emission on the red-shifted
side) the data, if extended to the shell center, would intersect 0 AU at approximately the same 
LSR velocity ($\sim$$+$13 km s$^{-1}$).  The extensions of the data for the different shells on 
the blue-shifted side would not intersect 0 AU close to the same velocity at all.  We note
that the velocity of $+$13 km s$^{-1}$ corresponds to the velocity in both the spectra in 
Figure 3 where the steepest part of the spectrum begins.  We also point out that full width at zero
power (FWZP) of both the SiS and HC$_3$N lines are essentially the same, despite the fact that
the spatial distribution and extent is much different.  This suggests that the gas is 
accelerated quickly in the inner region of the CSE to the expansion velocity.  The strong asymmetric
character (with the red-shifted side stronger) of both these transitions as well as others (e.g.\ 
the SiS, J $=$ 5$\rightarrow$4 transition; Lindqvist et al. 2000) must provide clues to detailed
kinematics of the gas.

While in general it is clear from Figure 1 that the HC$_3$N emission from RW~LMi consists 
mostly of broken rings (or shells) of emission, there are some velocity channels which, if
the emission is traced carefully, might be considered as a spiral shape (c.f.\ the velocity 
channel at $-$7.5 km s$^{-1}$ in Figure 1).  Dinh-V-Trung \& Lim suggested such a spiral for
the 45.5 GHz transition and there is at least one beautiful example of spiral {\bf dust} emission
in a CSE (AFGL 3068; Mauron \& Huggins 2006), but if there is such a spiral in our images 
it has the opposite handedness than that shown in the discussion of Dinh-V-Trung \& Lim.  Thus 
we think that the possibility of molecular spiral emission in RW~LMi is unlikely, though we 
do not completely rule it out.

Dinh-V-Trung \& Lim (2008) have used high angular resolution observations using the VLA to trace
several incomplete arcs (or shells) of HC$_3$N emission (at the 45.5 GHz transition) and HC$_5$N 
emission (transitions at 42.6 and 24.0 GHz) in the prototypical AGB carbon star IRC+10\arcdeg216,
similar to what we have observed in RW~LMi.  Dinh-V-Trung \& Lim (2008) estimate the kinematic
timescale between the shells is in the range of $\sim$120 to $\sim$360 yr.  This is also
similar to the timescales shown in Table 1, based on our data for RW~LMi.  The molecular arcs 
in IRC+10\arcdeg216 appear to coincide closely with the dust arcs shown by Mauron \& Huggins (2000).
For RW~LMi, Schmidt et al. (2002) have discovered several faint diffuse arcs of reflected light
using deep HST NICMOS imaging.  These arcs appear at radial distances of 1.0, 1.3, 3.8, and
4.5 arcseconds from the star, and are most prominent to the south-southwest of the star.  Only 
one of these reflected light arcs may be coincident with the HC$_3$N shell (at 4.5 arcseconds
radius) based on angular radius alone.  In addition, as can be seen in Figure 1, the most
prominent parts of the rings seen in HC$_3$N are typically not in the south-southwest part 
of the images.

\subsubsection{SiS Emission in RW~LMi and IK~Tau}

Figure 2 shows the SiS emission from both AGB stars averaged over the line.  We note that
there is a large difference in angular size of the SiS emission between the two stars.  This 
may be only a result of the generally weaker emission in IK~Tau, but the angular size difference 
is striking, and, assuming a distance to IK~Tau of 250 pc (Olofsson et al. 1998), the total 
physical extent of the SiS emission is different by a factor 4 (the emission in RW~LMi --- 
$\sim$2000 AU is 4 times larger than in IK~Tau --- $\sim$500 AU).  The peak of the SiS 
emission is coincident with that of the radio continuum emission which defines the stellar 
position.   

The size estimate for RW~LMi is consistent with the modelling of the SiS molecular
abundance by Sch\"oier et al. (2007).  The total SiS envelope size based on their Table 3
is $\sim$1700 AU.  The modelling is based on single-dish spectra; however, for RW~LMi the model 
appears to fit the interferometric data from the SiS, J $=$ 5$\rightarrow$4 transition at 
$\sim$90.7 GHz (Lindqvist et al. 2000) rather well.  For IK~Tau, the model gives a total SiS 
envelope size of 2200 AU (a factor 4 larger than our measurement).  Sch\"oier et al. point
out that their single-Gaussian abundance distribution gives a poor fit to the mm and submm-wave 
spectra of SiS for IK~Tau.  The addition of a compact, high abundance component to the model
improves the fit, especially for higher J transitions.  We speculate that the oxygen-rich
chemistry of IK~Tau may have some bearing on the detailed radial abundance of SiS in oxygen-rich
CSE.  It is clear that a larger sample of stars (both oxygen-rich and carbon-rich CSEs) observed with 
interferometric imaging at high angular resolution are needed to provide the best constraints
on such models.


\section{Summary}

We have used the ever-growing capabilities of the Expanded Very Large Array to perform a 
$\sim$1 GHz, targeted line survey of two AGB stars with $\sim$1 km s$^{-1}$ spectral resolution
(125 kHz at 18 - 50 GHz) and sub-arcsecond angular resolution.  The stars are: RW~LMi (with 
a carbon-rich CSE), and IK~Tau (with an oxygen-rich CSE). We find the following results in 
a preliminary analysis of the data:

$\bullet$ Continuum emission from both stars detected at 36.5 GHz is consistent with photospheric
emission from thermal gas.

$\bullet$ The HC$_3$N (J = 4$\rightarrow$3) emission from RW~LMi is bright enough to be
imaged in detail and shows strong asymmetries in the global line profile and partial
rings (shells) of emission.
We have provided a preliminary
analysis of the shell emission and find that the multiple shells can be traced from about 
400 years to 1200 years.  In addition, we have detected weak HC$_7$N (J $=$ 32$\rightarrow$31)
emission from RW~LMi.

$\bullet$ The SiS (J = 2$\rightarrow$1) emission from both stars is compact, compared with the 
HC$_3$N emission from RW~LMi.  The emission seems elongated for both stars rather than ring-like.
The extent of the detected SiS emission in IK~Tau is only one-quarter ($\sim$500 AU)
that of RW~LMi ($\sim$2000 AU) The SiS global line profile from RW~LMi is also strongly asymmetric
and in the same sense (red-shifted side stronger than blue-shifted side) as the HC$_3$N line.

By the end of the EVLA construction period (in early 2013) spectral imaging surveys should
be able to use 8 GHz per polarization with 125 kHz resolution (or better) in the WIDAR correlator 
simultaneously.  Thus the 3 receiver bands which span the 18 - 50 GHz range could be covered with 
5 frequency settings.  This will lead to a revolution in spectral imaging surveys of cm molecular 
transitions in AGB stars and related objects.

\acknowledgments
It is with sadness that we note the passing earlier this year of our colleague Fredrik Sch\"oier,
and we dedicate this short communication in his honor.

The National Radio Astronomy Observatory is a facility of the National Science
Foundation operated under cooperative agreement by Associated Universities, Inc.




{\it Facilities:} \facility{EVLA}.

\clearpage

\begin{table}
\begin{center}
\caption{Fitted and Derived Parameters of Possible HC$_3$N Shells in RW~LMi}
\begin{tabular}{ccccc}
\tableline\tableline
Shell & $v_{*}$ (km s$^{-1}$) & $r_{0}$ (AU) & $v_{exp}$ (km s$^{-1}$) & Age (yr) \\
\tableline
1 & -2.6 & 880 & 11.3 & 370 \\
2 & -3.1 & 2000& 12.0 & 800 \\
3 & -2.0 & 2500& 15.2 & 780 \\
4 & -1.2 & 3650& 14.8 & 1200 \\
\tableline
\end{tabular}
\end{center}
\end{table}

\clearpage

\begin{figure}
\epsscale{2.00}
\hspace {0.2in} \vspace{1.0in} \includegraphics[angle=0,scale=0.8] {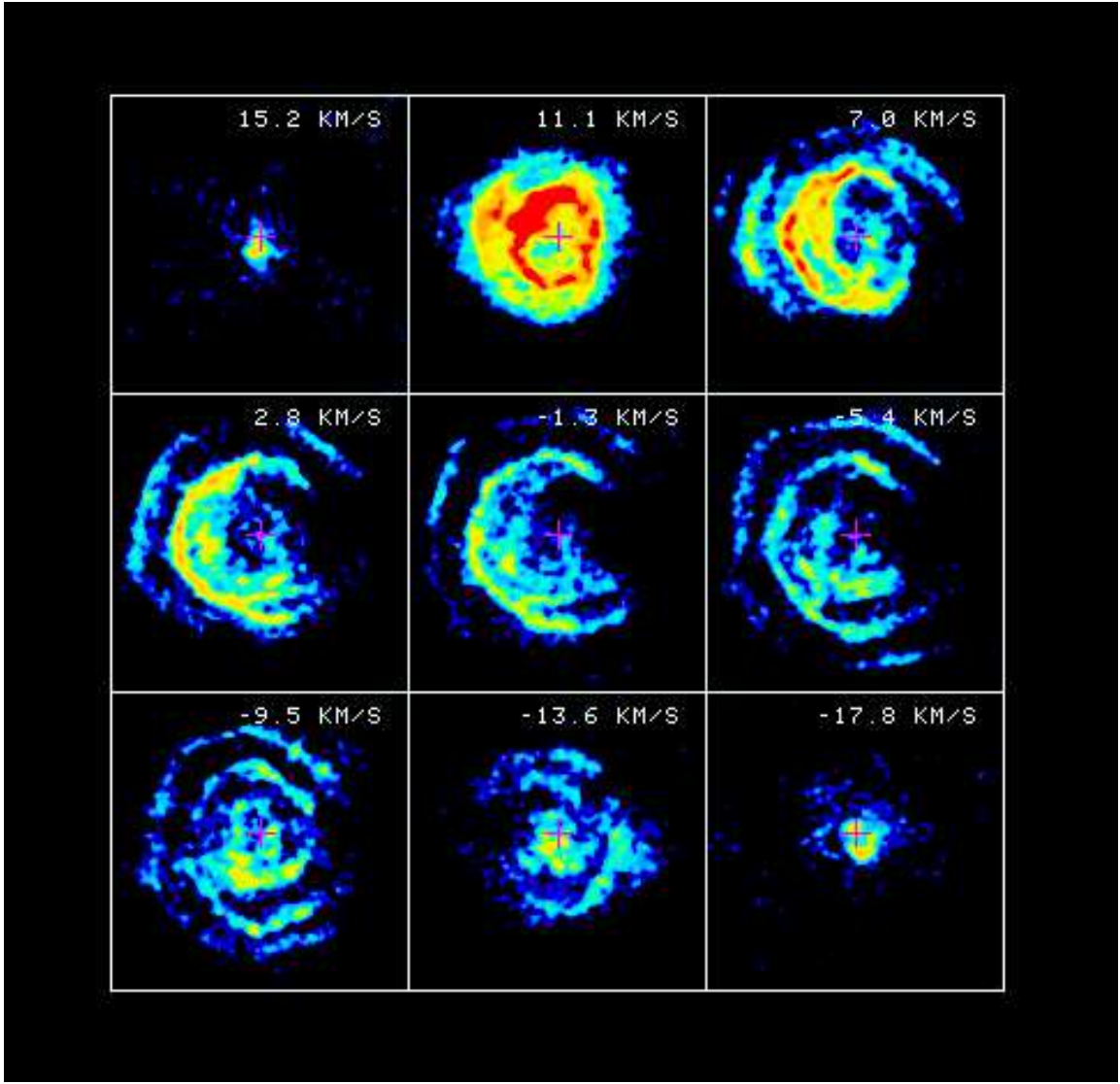}
\caption{Selected channel images of the HC$_3$N (J $=$ 4$\rightarrow$3, rest frequency
36392.332 MHz) emission from RW~LMi.  This selection 
spans the range of emission for the HC$_3$N transition, and each panel is every fourth
channel. Each image is $\sim$21 arcseconds on a side and is labeled with the velocity
with respect to the LSR.  The cross marks the position of the continuum source; the
size of the cross is about ten times the position error of the continuum measurement.
North is up and east is to the left. }
\end{figure}

\clearpage

\begin{figure}
\epsscale{1.00}
\plotone{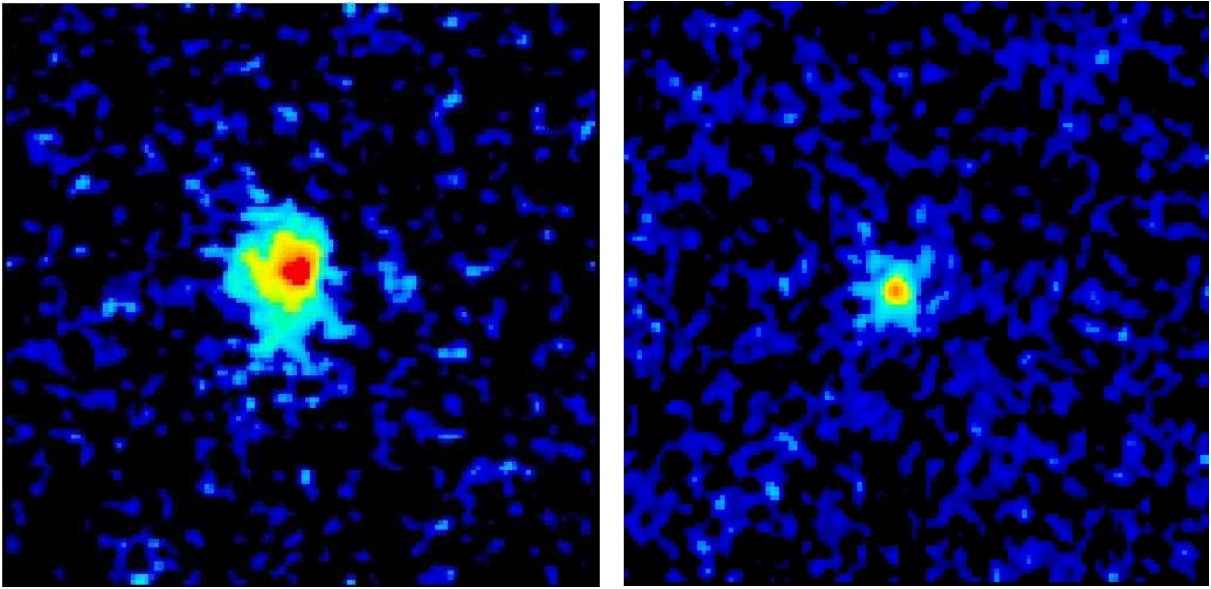}
\caption{The SiS (J$=$2$\rightarrow$1) emission for both sources, averaged over the emission
line, and after subtracting the continuum emission.  Each panel is $\sim$21 arcseconds on a side;
north is up and east is to the left.  Left:  the emission from RW~LMi.  Right: the emission from IK~Tau.
}
\end{figure}

\clearpage

\begin{figure}
\epsscale{1.00}
\plotone{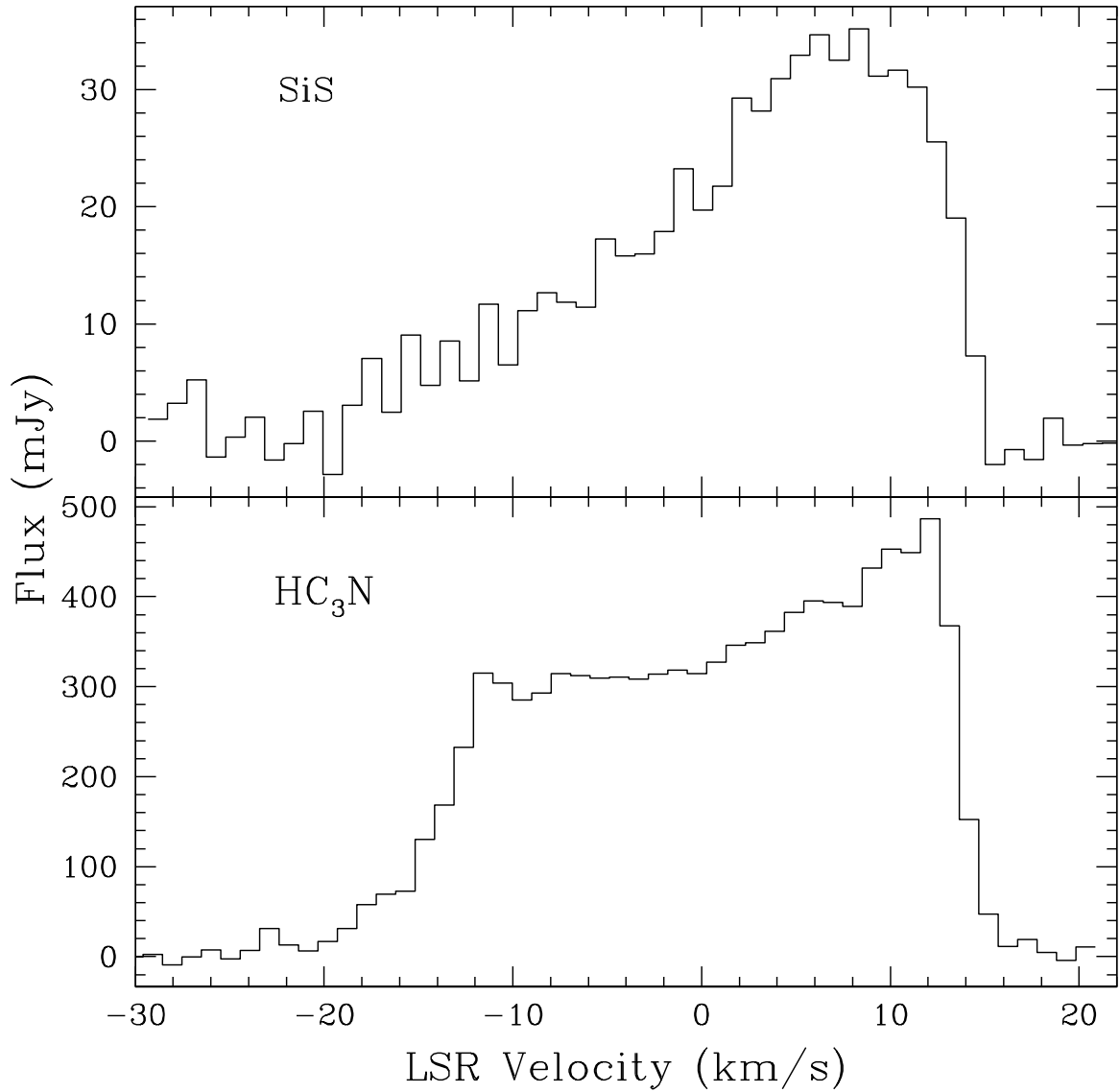}
\caption{Spectra of the HC$_3$N (J $=$ 4$\rightarrow$3) line (bottom) and the SiS (J $=$ 2$\rightarrow$1)
line (top) from RW~LMi.  The spectra were produced by summing the emission in a give rectangular region
for each spectral channel (the summed region is much smaller for the SiS line as compared with the 
HC$_3$N line).
}
\end{figure}

\clearpage

\begin{figure}
\epsscale{1.00}
\plotone {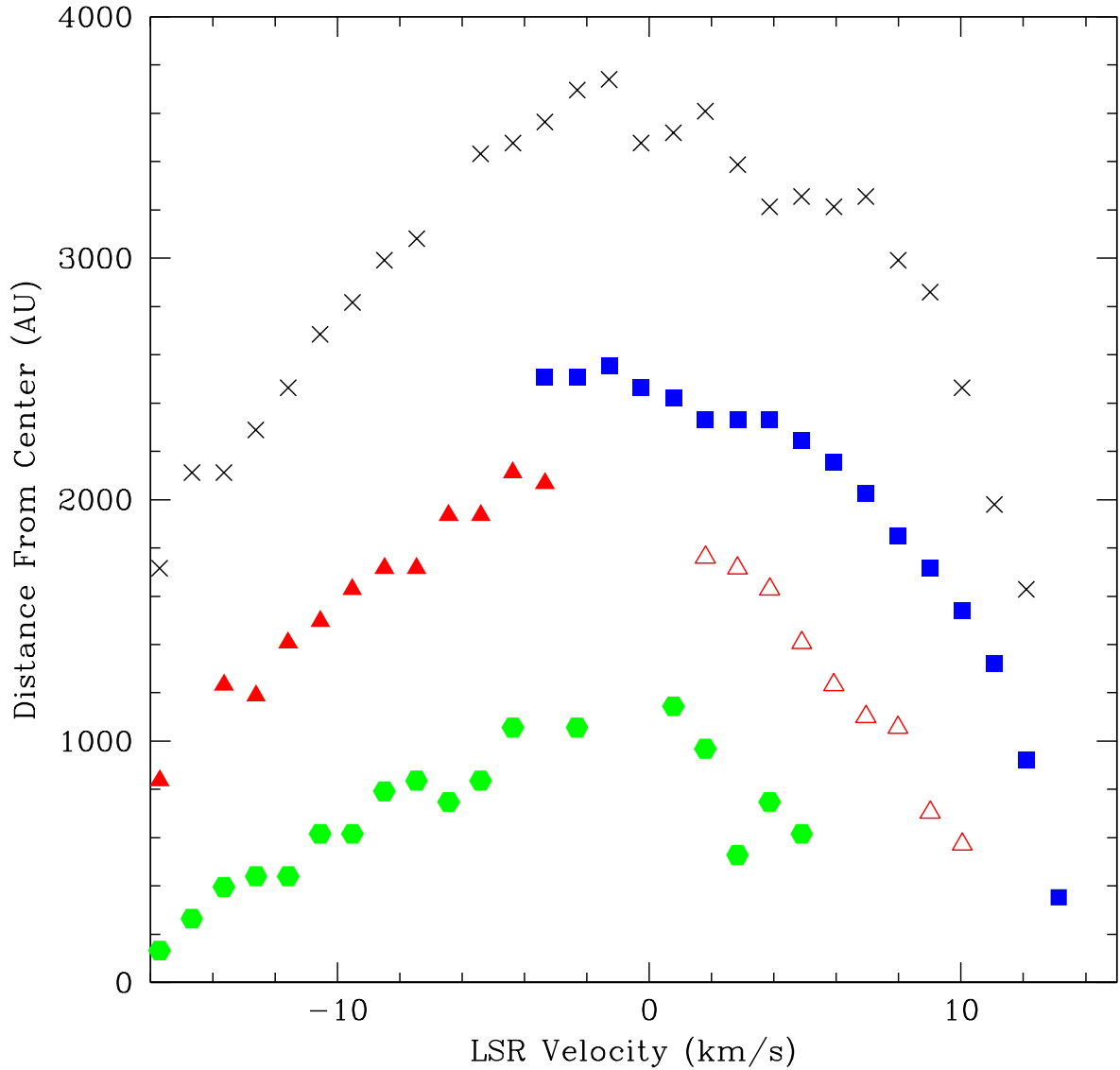}
\caption{Plots of the radius vs. LSR velocity for the shells of the HC$_3$N 
emission in RW~LMi, as determined using the IRING task of AIPS (which integrates the emission
in azimuthal rings for each channel). The errors in the distance of these rings are 
approximately the size of the symbols.  The different sections of the red triangles (one section
is solid; one is open) may or may not be the same ring, since it is difficult to determine, given 
the general asymmetric nature of the emission (see Figure 1 and the text).}
\end{figure}

\end{document}